\RequirePackage{amsmath}
\documentclass[smallextended]{svjour3}
\usepackage{graphics}
\usepackage{amssymb}
\usepackage{natbib}
\usepackage{bm}
\usepackage{xcolor}
\usepackage{hyperref}

\smartqed
\journalname{Exp Astron}

\begin{document}
\sloppy

\title{Attitude determination for nano-satellites -- I. Spherical projections for large field of view infrasensors}

\author{Korn\'el Kap\'as \and Tam\'as Boz\'oki \and \\ Gergely D\'alya \and J\'anos Tak\'atsy \and \\ L\'aszl\'o M\'esz\'aros \and Andr\'as P\'al}

\institute{%
K. Kap\'as, T. Boz\'oki, G. D\'alya, J. Tak\'atsy, L. M\'esz\'aros, A. P\'al
        \at Konkoly Observatory of the Research Centre for Astronomy and Earth Sciences, Budapest, Hungary \\
E-mail: kornel.kapas@ttk.elte.hu \\
K. Kap\'as, G. D\'alya, J. Tak\'atsy
        \at E\"otv\"os Lor\'and University, Institute of Physics, P\'azm\'any P\'eter stny. 1/A, Budapest H-1117, Hungary\\
T. Boz\'oki
        \at Institute of Earth Physics and Space Science (ELKH EPSS), Csatkai Endre utca 6-8, Sopron H-9400, Hungary\\
T. Boz\'oki
        \at Doctoral School of Environmental Sciences, University of Szeged, Aradi v\'ertan\'uk tere 1, Szeged H-6720, Hungary \\
J. Tak\'atsy
        \at Institute for Particle and Nuclear Physics, Wigner Research Centre for Physics, Konkoly-Thege Mikl\'os \'ut 29-33, Budapest H-1121, Hungary
}

\maketitle

\begin{abstract}
Due to the advancement of nano-satellite technology, CubeSats and fleets of CubeSats can form an alternative to high-cost large-size satellite missions with the advantage of extended spatial coverage. One of these initiatives is the Cubesats Applied for MEasuring and LOcalising Transients (CAMELOT) mission concept, aimed at detecting and localizing gamma-ray bursts with an efficiency and accuracy comparable to large gamma-ray space observatories. While precise attitude control is not necessary for such a mission, attitude determination is an important issue in the interpretation of scintillator detector data as well as optimizing downlink telemetry. The employment of star trackers is not always a viable option for such small satellites, hence another alternative is necessary.

A new method is proposed in this series of papers, utilizing thermal imaging sensors to provide simultaneous measurement of the attitude of the Sun and the horizon by employing a homogeneous array of such detectors. The combination with Sun and horizon detection w.r.t. the spacecraft would allow the full 3-DoF recovery of its attitude. In this paper we determine the spherical projection function of the MLX90640 infrasensors planned to be used for this purpose. We apply a polynomial transformation with radial corrections to map the spatial coordinates to the sensor plane. With the determined  projection function the location of an infrared point source can be determined with an accuracy of $\sim40^\prime$, well below the design goals of a nano-satellite designed for gamma-ray detection. 
\end{abstract}

\section{Introduction}
\label{sec:introduction}

Nowadays the increasing importance of small satellite missions in space science and technology is obvious. A fleet of CubeSats provide a viable alternative for a unique, high-cost and large satellite, where the former one usually implies lower funding requirements and the advantage of extended spatial coverage, in accordance with the orbital distribution. Many such missions are now underway all the way from microsatellite based internet providers through the detection of exoplanets to X-ray measurements of the halo of the Milky Way (e.g. Halosat, \citealt{jahoda2019}; ASTERIA, \citealt{smith2018}).

Cubesats Applied for MEasuring and LOcalising Transients (CAMELOT, \citealt{werner2018,pal2018,ohno2018}) is such a nano-satellite fleet mission concept aimed at detecting gamma-ray bursts with a nearly full spatial coverage of the sky. An individual CAMELOT satellite design \citep{werner2018} is based on 4 panels of $150\times 75\times 5\,{\rm mm}$ caesium-iodine (CsI) scintillator detectors while each of these detector panels are monitored with a set of multi-pixel photon counters (MPPCs). Localization of the gamma-ray bursts are then based on the triangulation of the events provided by precise GPS-based time-stamping on multiple satellites \citep{pal2018} which provides a detection accuracy in the order of ten minutes of arc \citep{ohno2018}. Since this combination of MPPCs with scintillators has never flown before, at the present stage of the project our goal is to design and build a 1U satellite ($100\times 100\times 113.5\,{\rm mm}$) matching the CubeSat standards for testing many of the scientific and electronic configurations of the later 3U CAMELOT satellites, where the gamma-ray detection is provided by a single panel of CsI crystal, having a size of $75\times 75\times 5\,{\rm mm}$.

One of the difficulties connected to nano-satellites is the accurate determination of their orientation. On large-size satellites the orientation is usually determined by costly, large-size star trackers, based on the apparent brightness and angular separation of numerous stars in the field of view. As these systems form an individual unit within the satellite they do not fit the small size and/or power budget criterion of nano-satellites. Therefore we propose a new, cost-efficient approach to this problem namely the utilization of thermal imaging sensors in order to determine the direction of the Sun \emph{and} the horizon with respect to the satellite with sub-degree accuracy. 

\begin{figure}
\begin{center}
\resizebox{35mm}{!}{\includegraphics{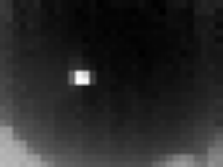}}\hspace*{3mm}%
\resizebox{35mm}{!}{\includegraphics{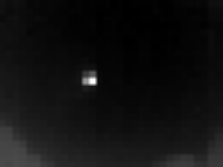}}\hspace*{3mm}%
\resizebox{35mm}{!}{\includegraphics{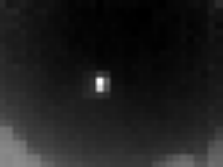}}
\end{center}
\caption{Some example thermal images of the sky, taken by a fixed MLX90640
sensor as the Sun moves in front of it with a synodic apparent motion. The
frames are separated by $\sim 17$ minutes in time. While the horizon is
not covered fully in this setup, it can clearly be seen at the bottom
of these images.}
\label{fig:sun}
\end{figure}

\begin{figure}
\begin{center}
\resizebox{90mm}{!}{\includegraphics{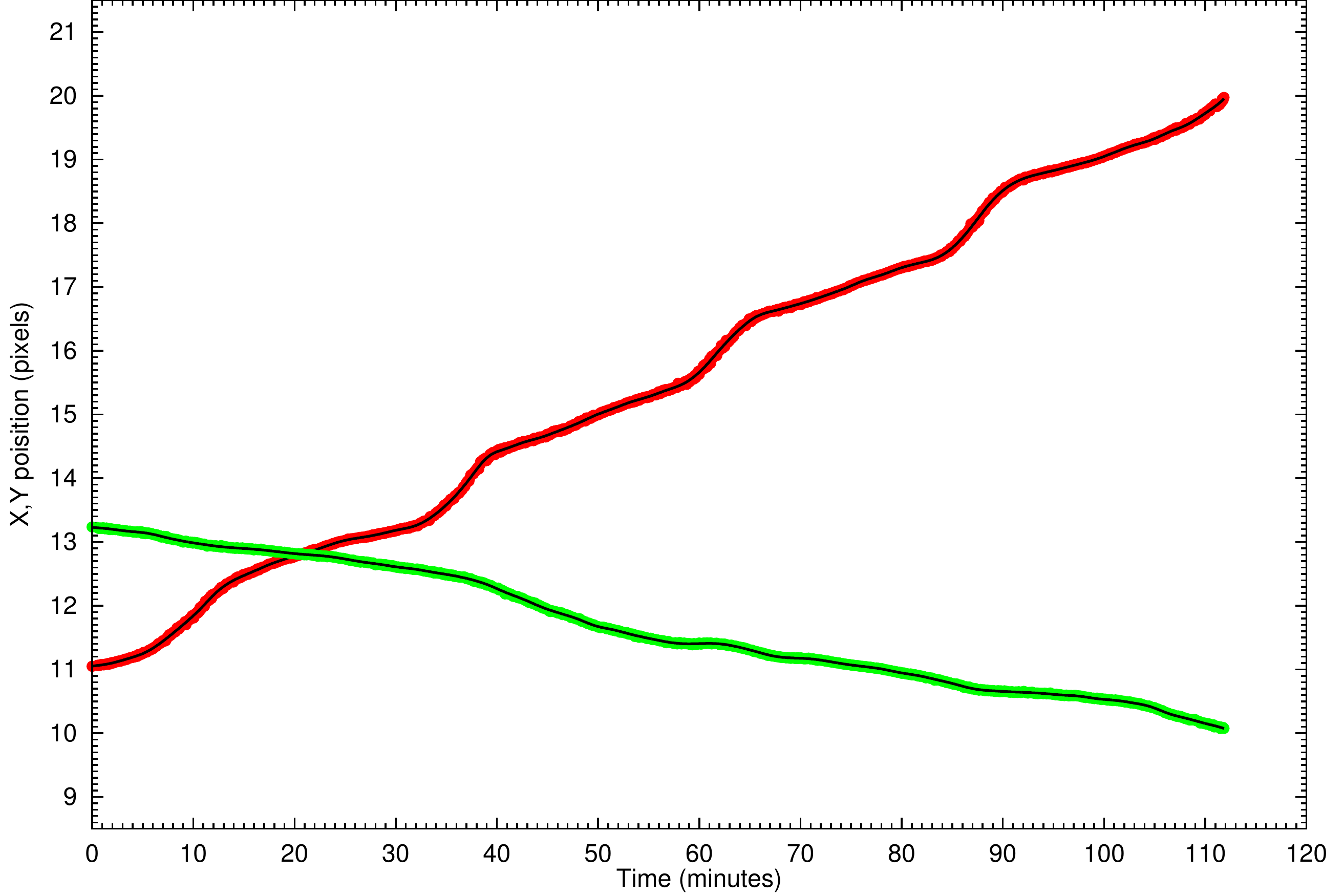}}%
\end{center}
\caption{Absolute pixel coordinates of the best-fitted centroid
position corresponding
to the Sun as seen by an MLX90640 sensors during a run of roughly 110 minutes.
In this ground-based experiment, the sensor was fixed while the apparent
motion of the Sun reflected the rotation of Earth. The red curve
corresponds to the $x$ coordinate while the green one is for the $y$ coordinate.
The best-fitted spline curves are superimposed as black. The residual is in
the order of few hundreds of a pixel. }
\label{fig:absxy}
\end{figure}

\begin{figure}
\begin{center}
\resizebox{95mm}{!}{\includegraphics{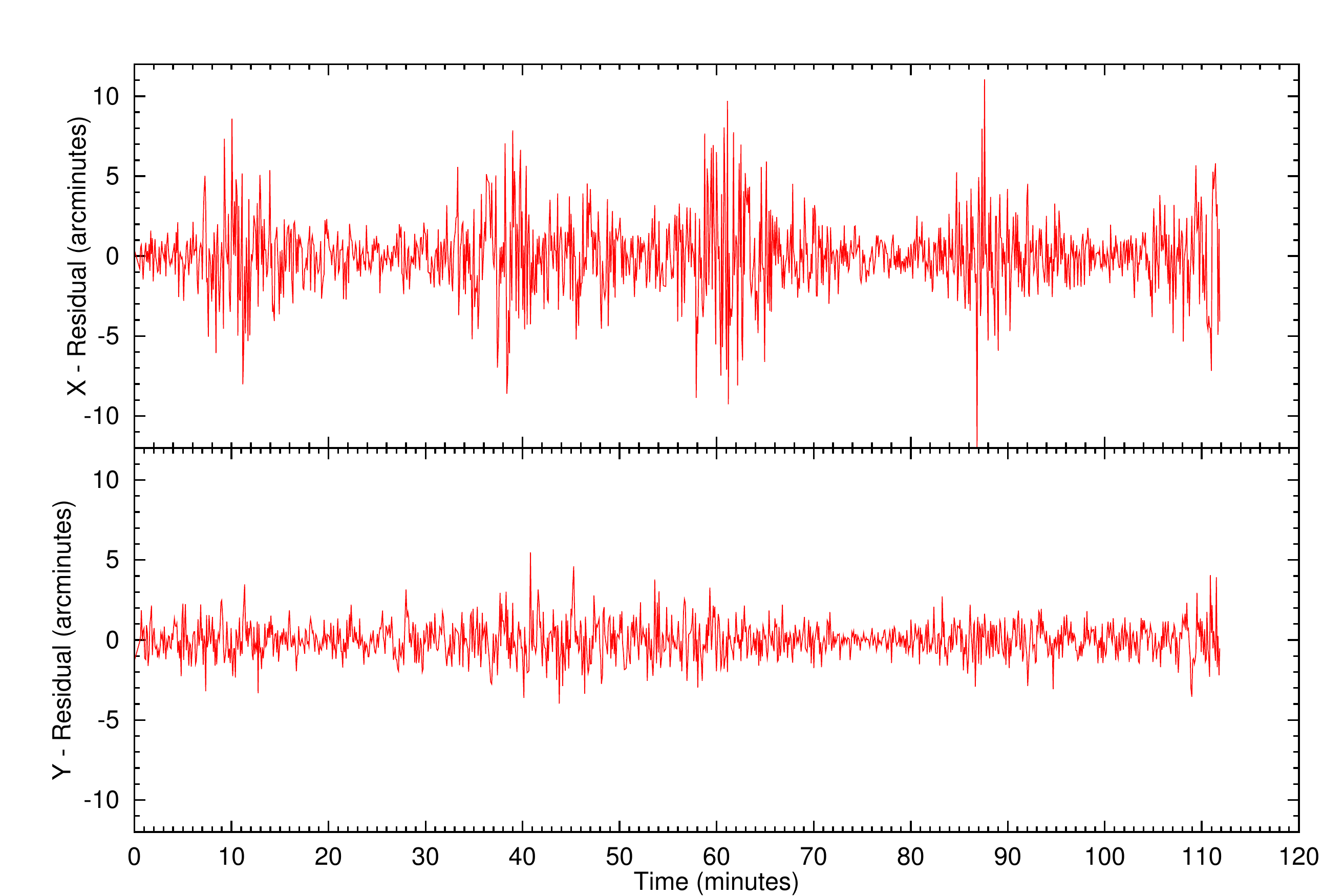}}%
\end{center}
\caption{The $x$ and $y$ residuals of the spline fitting procedure
corresponding to the experiment shown in Fig.~\ref{fig:absxy}. The typical
residuals in the order of a few hundreds of a pixel. This is equivalent
to a few arcminutes on the sky or $\sim5\cdot10^{-4}$ radians.}
\label{fig:resxy}
\end{figure}

We have carried out preliminary tests where the apparent motion of the Sun was monitored by an MLX90640 sensor from a fixed position. Some example thermal images are shown in Fig.~\ref{fig:sun}. In Fig.~\ref{fig:absxy} the best-fitted centroid positions of the Sun are displayed as a function of time. By removing the tendency of the curves (represented by spline fits) we obtain the residuals shown in Fig.~\ref{fig:resxy}. As a conclusion we found that the \emph{precision} of determining the position of a point-like source might be as low as a few minutes of arc, however, it does not imply that the overall accuracy of such a setup will be comparable to this precision level.

As we will see later on (see Sec.~\ref{sec:setup}), 6 pieces of the proposed sensor type would allow us to provide a full (i.e. $4\pi$ steradian) spatial coverage including Earth (where its obscuring depends on the altitude of the satellite orbit) and the Sun, when the satellite is not in the shadow of Earth. Partial attitude reconstruction is still possible when the Sun is obscured and/or some of the detectors cannot provide a thermal image (due to, e.g. permanent or temporary failure). In this case, dead reckoning can be involved as a fallback where the (partial) attitude information provided by the imaging sensors are evaluated in coincidence with data provided by MEMS gyroscopes (cf. exploiting suitable Kalman filters). The accurate determination of the attitude is also crucial for enhancing the efficiency of the satellite telemetry. Furthermore, attitude determination will serve complementary in source localization to triangulation using simultaneous gamma-ray measurements of multiple satellites.

Consequently, one of the most essential steps towards attitude determination with thermal imaging sensors is the derivation of the function that quantifies how a point source (such as the Sun) is mapped by the sensors to its pixel plane. Due to the large field of view and the manufacturing process of these sensors, this mapping contains a significant amount of distortion which has both radial components and anisotropic terms. In the present paper we propose an experimental method for extracting this spherical mapping information and the coefficients of the related projection function (i.e. the function connecting the unit vector pointing from the sensor towards the source to the pixel coordinates of the source on the IR image). In Sec.~\ref{sec:setup} we describe our experimental setup and the measurements we have carried out. We evaluate the results in Sec.~\ref{sec:Res} and draw our conclusions in Sec.~\ref{sec:conclusions}. Once these functions are well characterized, one can proceed with the recovery of the full attitude information and the integration of gyroscope data for better long-term accuracy in the case of partial attitude knowledge. This topic is going to be covered in an upcoming paper.

\section{Experimental setup}
\label{sec:setup}

\begin{figure}
\begin{center}
\resizebox{105mm}{!}{\includegraphics{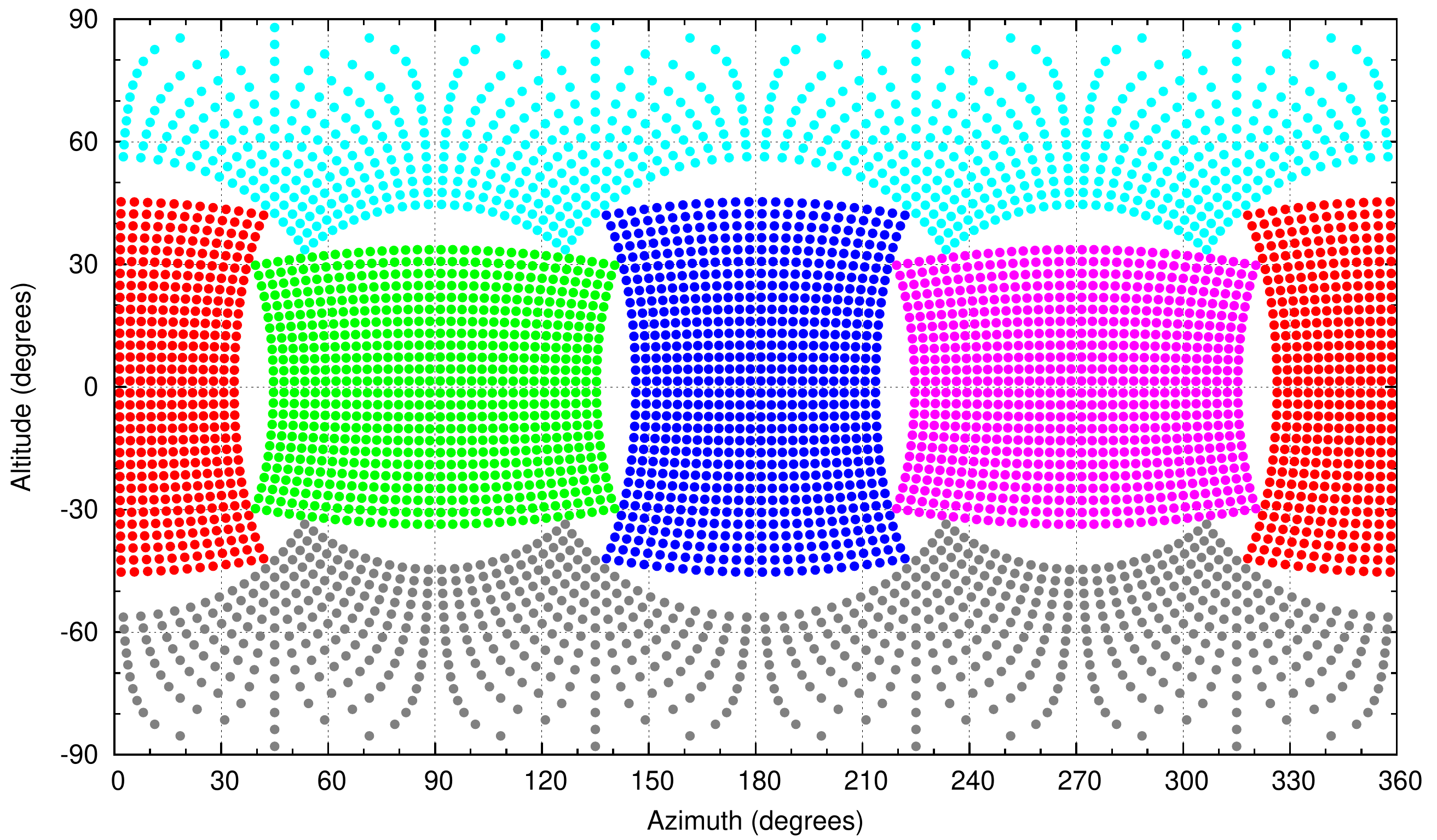}}%
\end{center}

\begin{center}
\resizebox{75mm}{!}{\includegraphics{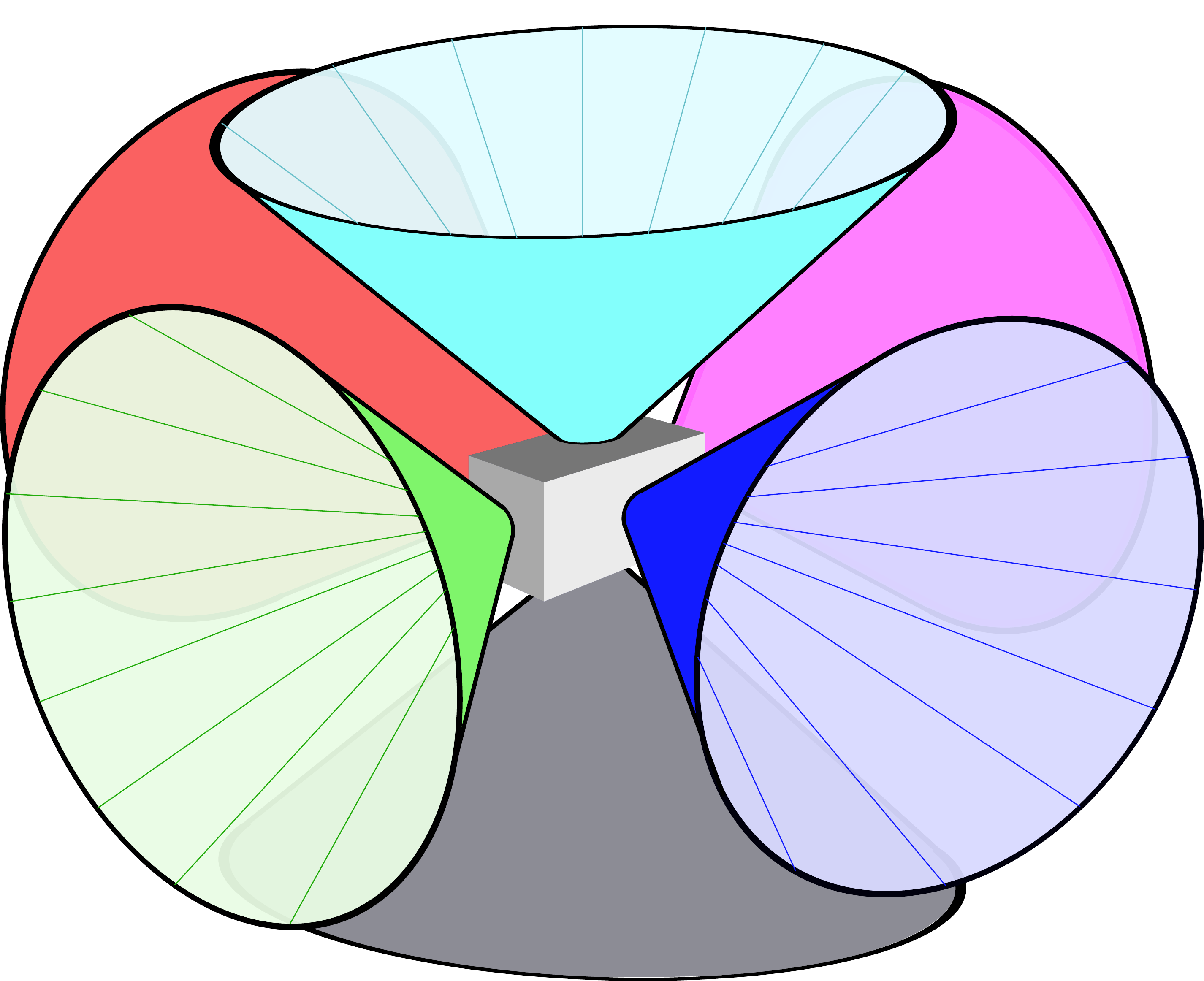}}%
\end{center}

\caption{Full-sky projection of six perpendicular MLX90640 sensors and
the corresponding pixel centroid coordinates in an altitude-azimuth
coordinate system (top), and an illustration of the satellite together with the viewing cones of infrasensors installed on each side of the satellite (bottom). The colored pixel centroid coordinates correspond to colored cones with the same color.}
\label{fig:fullsky}
\end{figure}

MLX90640 is a small-size, low-cost IR sensor having 32$\times$24 pixels and a relatively large, 110$\times$75 degree field of view \citep{MLX}. This coverage by a single sensor implies that six of these sensors, placed on the six sides of a cube, could cover the full sphere, see Figure \ref{fig:fullsky}. Note that the attitude of the sensors needs to be aligned accordingly due to the rectangular field of view. This property makes these sensors adequate to employ in CubeSat designs. 

\begin{figure}[!ht]
\centering
\resizebox{!}{56mm}{\includegraphics{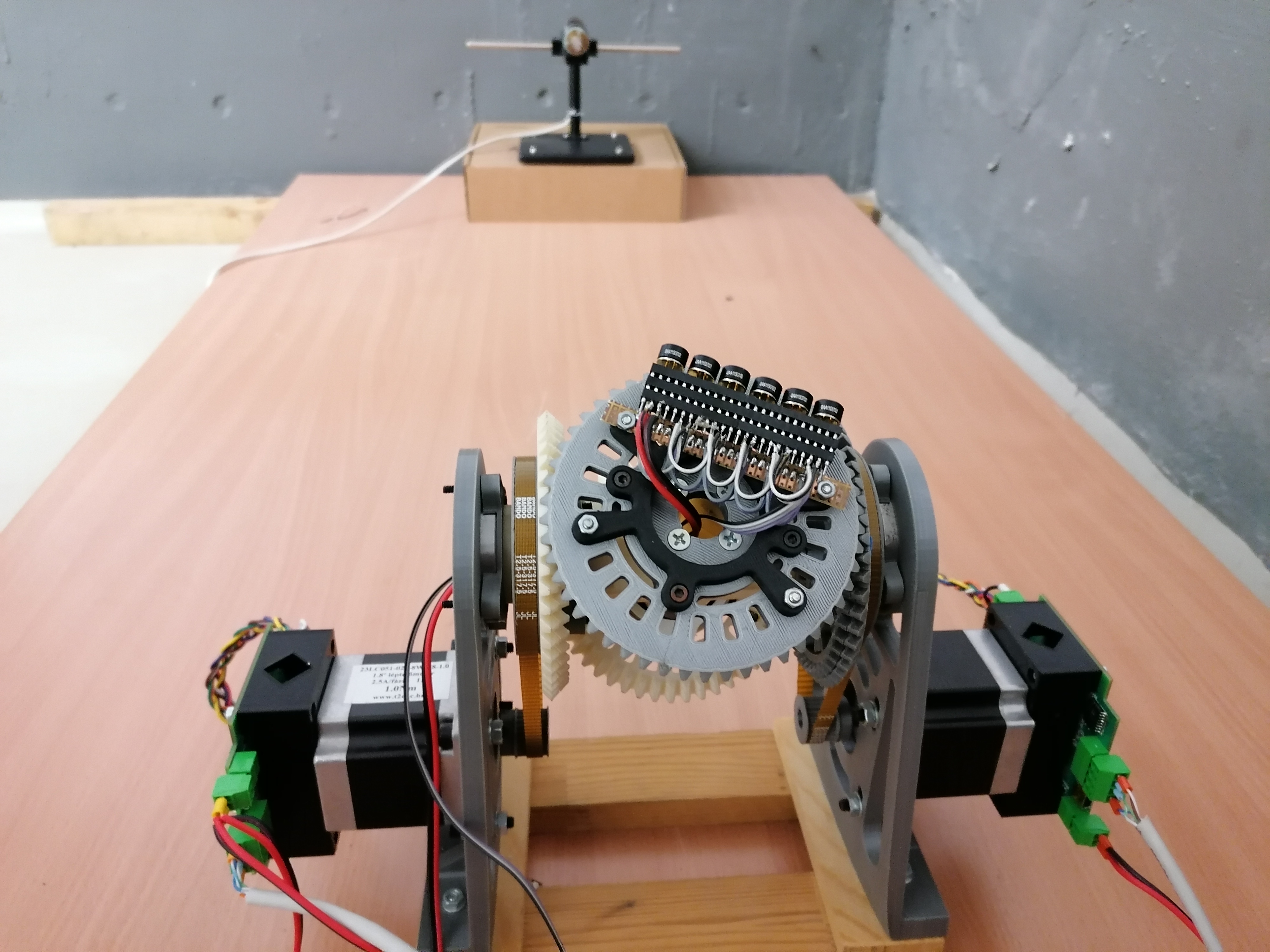}}
\resizebox{!}{56mm}{\includegraphics{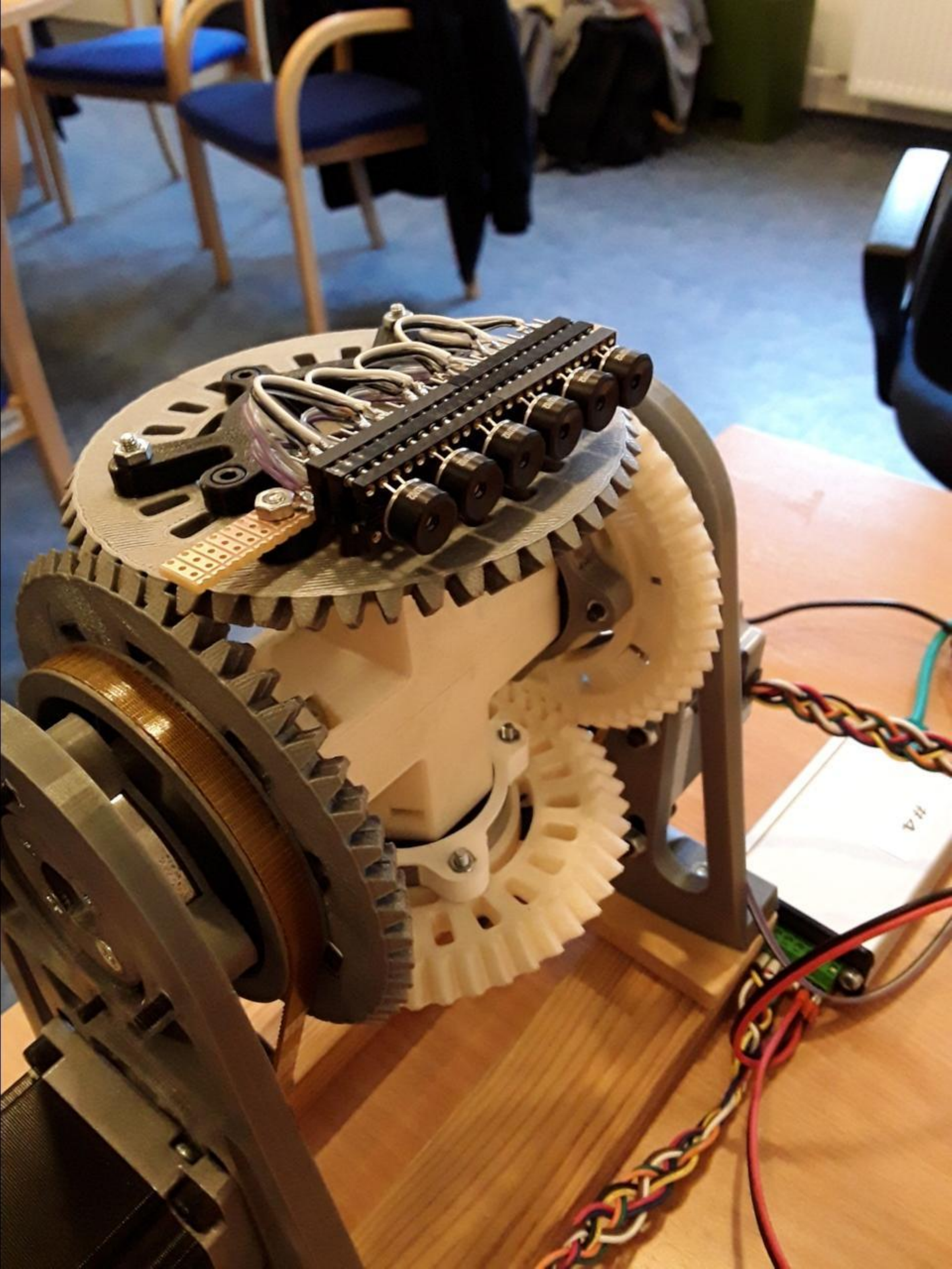}}

\begin{center}
\resizebox{105mm}{!}{\includegraphics{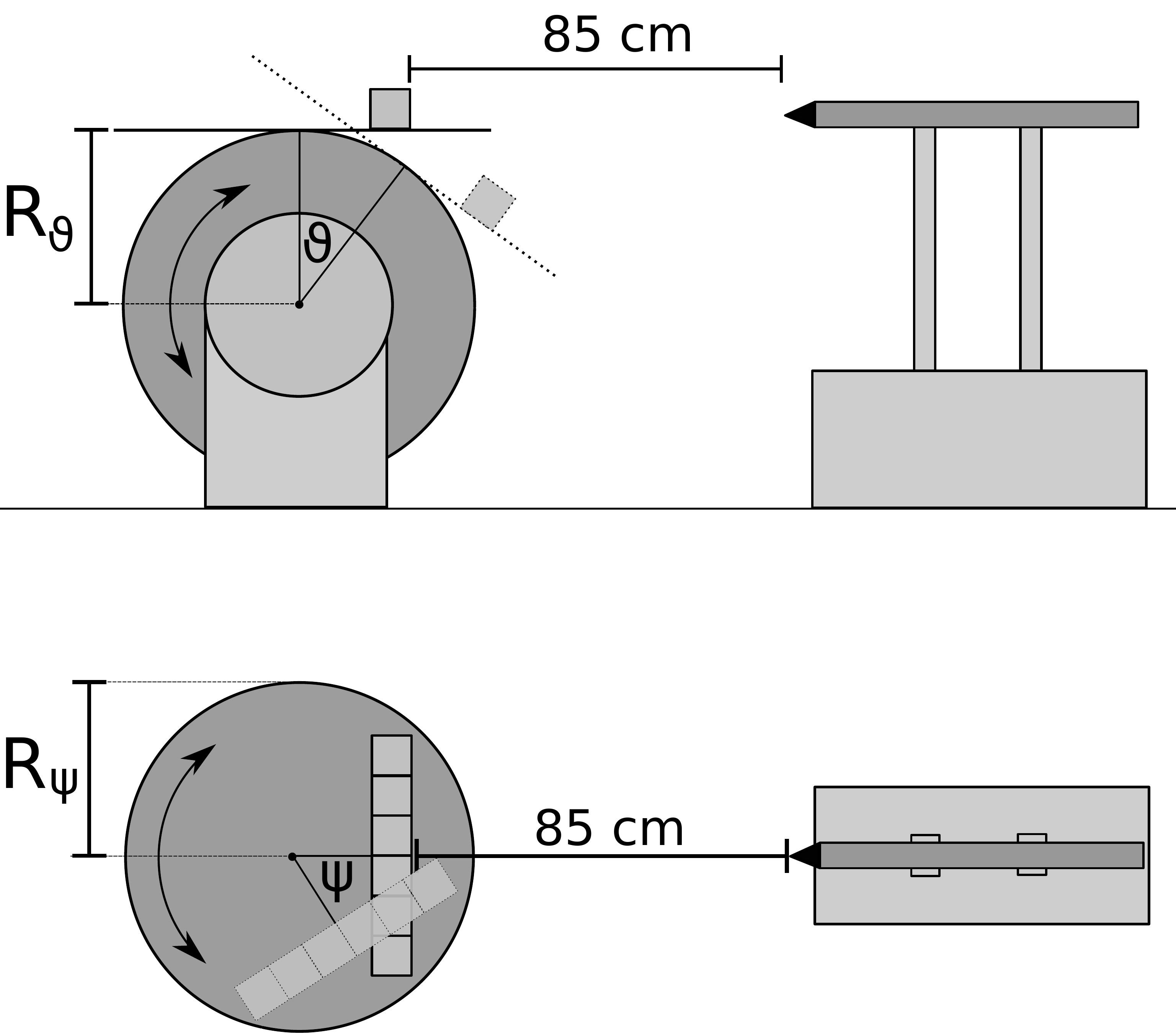}}%
\end{center}

\caption{The experimental setup. The top images show the full setup with the soldering gun in the back on the second basement floor of the Konkoly Observatory and the six detectors from a closer view, while the illustrations in the bottom are schematic drawings of the experimental setup viewed from the side and from top with the pitch ($\vartheta$) and yaw ($\psi$) angles also indicated.}
\label{pic:berendezes}
\end{figure}

In order to determine the projection function of such a sensor we have built a device that is able to set its orientation in a predefined way with respect to a fixed IR source. The device we have used for this purpose was built in a similar fashion to that used for the attitude calibration of MEMS accelerometers \citep{meszaros2014}. It consists of four bevel gears (see Fig. \ref{pic:berendezes}) from which the two vertical ones are  driven by two independent stepper motors. On the top of the device, six MLX90640 infrasensors have been placed and connected via an I$^2$C bus and 3.3 V power supply. In that way the sensors can be moved on a spherical surface and their positions can be characterized by two angles: the pitch ($\vartheta$) and the yaw ($\psi$) of the system. 

In its $(\vartheta,\psi)=(0,0)$ position the device points towards a soldering iron located $\sim85\,{\rm cm}$ from the sensors, serving as the point-like IR source in our experiments. The use of a soldering iron with a diameter of 8 mm as a reference source is justified by the similar angular size of the soldering iron from $85$ cm and the Sun from a low Earth orbit, as well as by the fact that the thermal signal in the sensor sensitivity regime yielded by the soldering iron at this distance and wavelength is comparable to the radiation flux of the Sun. This is mainly due to the overlapping of the Rayleigh-Jeans tail of the solar spectrum and the sensitivity regime of the MLX90640 sensors: while the solar flux is well below of its peak, the soldering iron has the peak of its own black body radiation at this domain. The full setup of the experiment (including the soldering iron, the attitude control mechanism with the sensors and additional control computers) has been placed in a basement room ensuring spatially and temporally homogeneous temperature background for the measurements. The whole setup is remotely controlled, i.e. the process of moving the sensors as well as data collection can easily be automatized.

The pitch and yaw angles were selected from the ranges $\vartheta\in[-40^\circ,40^\circ]$ and $\psi\in[-50^\circ,50^\circ]$, respectively, in steps of $10^\circ$ in both directions. Choosing a position beyond these ranges would have resulted in the IR source being outside the field of view of the sensors. The sequence of $(\vartheta,\psi)$ pairs during the measurements was chosen randomly and multiple images were taken in a single position enabling us to determine the errors of the projection function parameters (see Sec. \ref{ssec:Map_func}). This process was then repeated for each sensor, as there could be manufacturing differences between them.

\section{Results}
\label{sec:Res}

\subsection{Evaluation of source positions from the IR images}
We utilised the FITSH package\footnote{https://fitsh.net/} \citep{pal2012}, an open-source software collection aimed at astronomical image and data processing, for evaluating the exact position of the IR source in pixel coordinates $(X,Y)$ automatically. We used the \texttt{fistar} and \texttt{fiphot} tasks, respectively, to have an initial estimation on all the possible source positions and then determine their locations more accurately. Usually, this algorithm identifies more than one source on the image, therefore we filtered them first by a rough estimate on the expected position of the IR source (based on a linear $X$($\vartheta$) and $Y$($\psi$) regression model) and then selected the source with the largest flux.

\begin{figure}[!ht]
\centering
\resizebox{!}{49mm}{\includegraphics{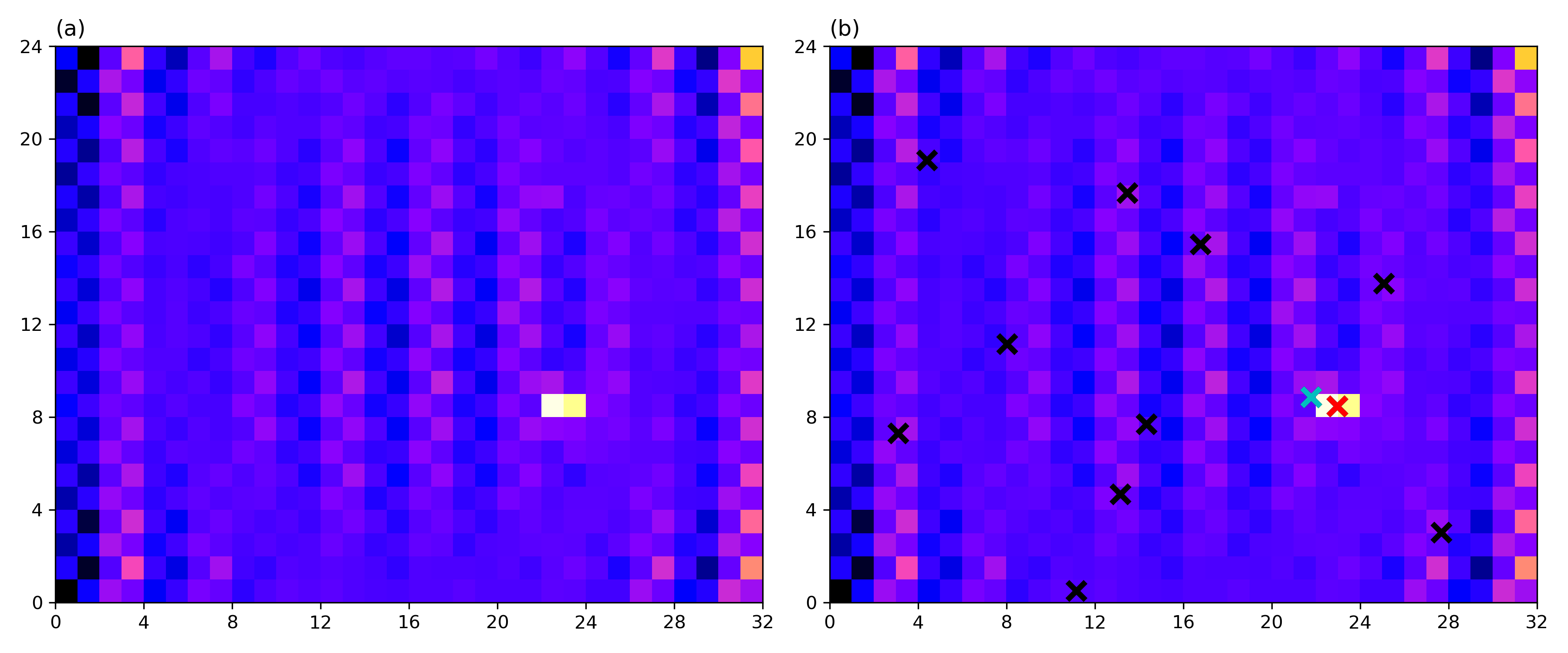}}
\caption{Evaluation of source positions: (a) the raw image and (b) all the extracted peak positions and the identified soldering gun position marked with black and red crosses, respectively. The blue cross marks a peak that matches the regression criterion but has a smaller flux compared to the red one.}
\label{pic:source_pos}
\end{figure}

The process is illustrated in Fig \ref{pic:source_pos}. On the raw image (Fig \ref{pic:source_pos}/a) the program first finds the peaks marked with black crosses (Fig \ref{pic:source_pos}/b), applies the regression criterion and then selects the peak with the largest flux near the expected source location (red cross in Fig. \ref{pic:source_pos}/b). The blue cross marks a peak that matches the regression criterion but has a smaller flux compared to the red one.

\subsection{Determining the projection function}
\label{ssec:Map_func}

Let $\mathbf{n}=(x,y,z)$ be the unit vector pointing to the direction of the soldering gun from the sensor and let $\Tilde{x}$ be the distance of a given sensor from the center of the sensor plane. The location of the sensor plane's center with respect to the pivot point of the motor, as a function of the pitch and yaw angles is:
\begin{equation}
    \mathbf{r}_{\Tilde{x}}(\vartheta, \psi) =
    \begin{pmatrix}
    -R_{\psi} \sin \psi + \Tilde{x} \cos \psi \\
    R_{\vartheta} \cos \vartheta + (R_{\psi} \cos \psi + \Tilde{x} \sin \psi) \sin \vartheta \\
    -R_{\vartheta} \sin \vartheta + (R_{\psi} \cos \psi + \Tilde{x} \sin \psi) \cos \vartheta
\end{pmatrix},
\end{equation}
where $R_{\vartheta}$ and $R_{\psi}$ are the radii of the bevel gears rotating the setup around the pitch and yaw axes, respectively.

\begin{figure}[!ht]
    \centering
    \resizebox{!}{42mm}{\includegraphics{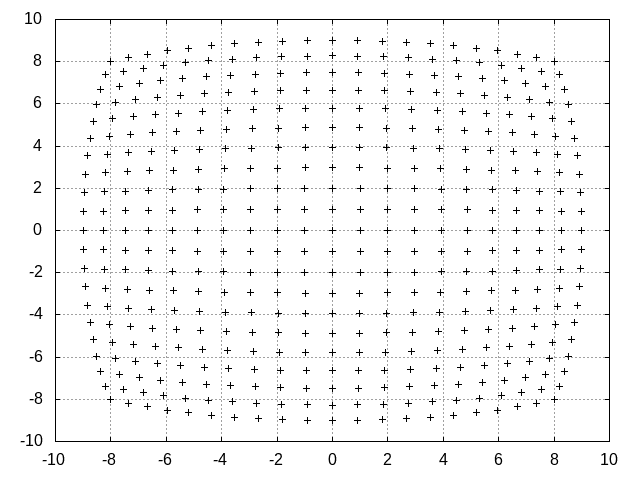}}
    \resizebox{!}{42mm}{\includegraphics{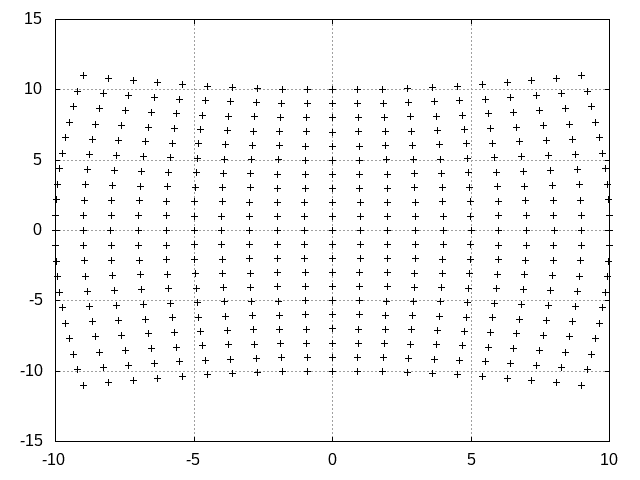}}
    \caption{Effect of the cubic distortions on a square lattice. Radial ($K_1$) on the left, anisotropic ($a_{12}$) on the right.}
    \label{fig:distortions}
\end{figure}

By denoting the location of the soldering gun in the frame centered on the midpoint of the motor by $\mathbf{P}_{\mathrm{m}}$ the formula connecting the pitch and yaw angles to the $(x,y,z)$ coordinates is the following:
\begin{equation}
    \mathbf{n}=\frac{\mathbf{r}}{|\mathbf{r}|} \:\: , \quad \quad \mathbf{r}=\mathbf{P}_{\mathrm{m}}-\mathbf{r}_{\Tilde{x}}(\vartheta, \psi)
\end{equation}
We would like to determine the function mapping these coordinates to the pixel coordinates $(X,Y)$ of the soldering gun on the image. For this we apply the following model. First, we apply a small, 3D rotation before using the projection function, to account for the ambiguities in the orientation of the individual sensors relative to the sensor holder:
\begin{equation}
\begin{pmatrix}
    x' \\
    y' \\
    z'
\end{pmatrix}
=
\begin{pmatrix}
    1 & \alpha & -\beta \\
    -\alpha & 1 & \gamma \\
    \beta & -\gamma & 1
\end{pmatrix} \cdot
\begin{pmatrix}
    x \\
    y \\
    z
\end{pmatrix}
=
\begin{pmatrix}
    x + \alpha y - \beta z \\
    -\alpha x + y + \gamma z \\
    \beta x - \gamma y + z
\end{pmatrix} \quad .
\end{equation}
The $\alpha$, $\beta$ and $\gamma$ angles are also determined together with the other parameters of the projection function. Then we introduce the tangential variables:
\begin{equation}
\eta=x'/z' \quad \quad \xi=y'/z'
\end{equation}
We use a projection function that has constant and linear terms with radial corrections, as well as higher-order polynomials. The radial corrections have the form:
\begin{equation}
    \eta'=\eta[1+K_1(\xi^2+\eta^2)+K_2(\xi^2+\eta^2)^2+...]\\
\end{equation}
\begin{equation}
    \xi' =\xi [1+K_1(\xi^2+\eta^2)+K_2(\xi^2+\eta^2)^2+...]
\end{equation}
We only keep the $K_1$ term, since higher-order terms are redundant in our case. Using these parameters the projection function can be written in the form:
\begin{equation}
    X_\mathrm{model}=a_{00}+a_{10}\,\eta'+a_{01}\,\xi'+a_{20}\,\eta^2+a_{11}\,\xi\eta+a_{02}\,\xi^2 +...
\end{equation}
\begin{equation}
    Y_\mathrm{model}=b_{00}+b_{01}\,\xi'+b_{10}\,\eta'+b_{02}\,\xi^2+b_{11}\,\xi\eta+b_{20}\,\eta^2 +...
\end{equation}
Due to the axial reflection symmetry of the optical system with respect to the $X$ and $Y$ axes, we omit all the terms with even powers of the variables (except the zero-order term). The linear cross-terms are also omitted, since the antisymmetric part is already included in the small rotation, and the symmetric part breaks the reflection symmetry. From the cubic terms we only keep the ones with the $a_{12}$ and $b_{21}$ coefficients and set $b_{21}=-a_{12}$. All higher-order terms are neglected. Hence, we are left with 9 independent parameters: $\alpha,\beta,\gamma,K_1,a_{00},b_{00},a_{10},b_{01},a_{12}$. The effects of the $K_1$ radial and the $a_{12}$ part of the projetion function are shown in Fig \ref{fig:distortions}.

\begin{figure}[!ht]
\centering
\resizebox{!}{80mm}{\includegraphics{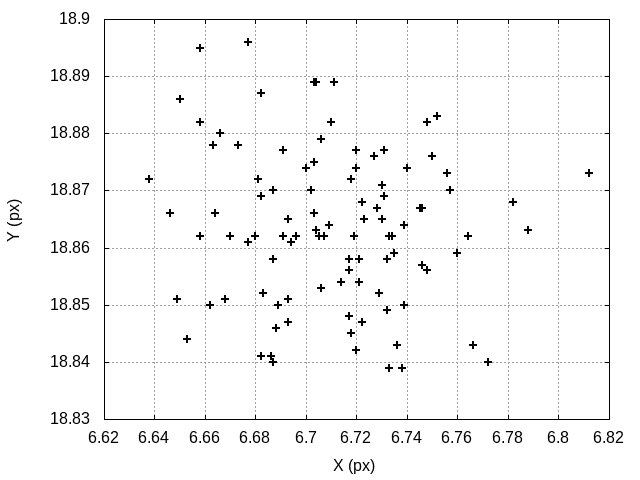}}
\caption{The scatter of the 100 found peak for a single sensor at fixed pitch and yaw values.}
\label{pic:peaks}
\end{figure}

\begin{figure}[!ht]
    \centering
    \resizebox{!}{80mm}{\includegraphics{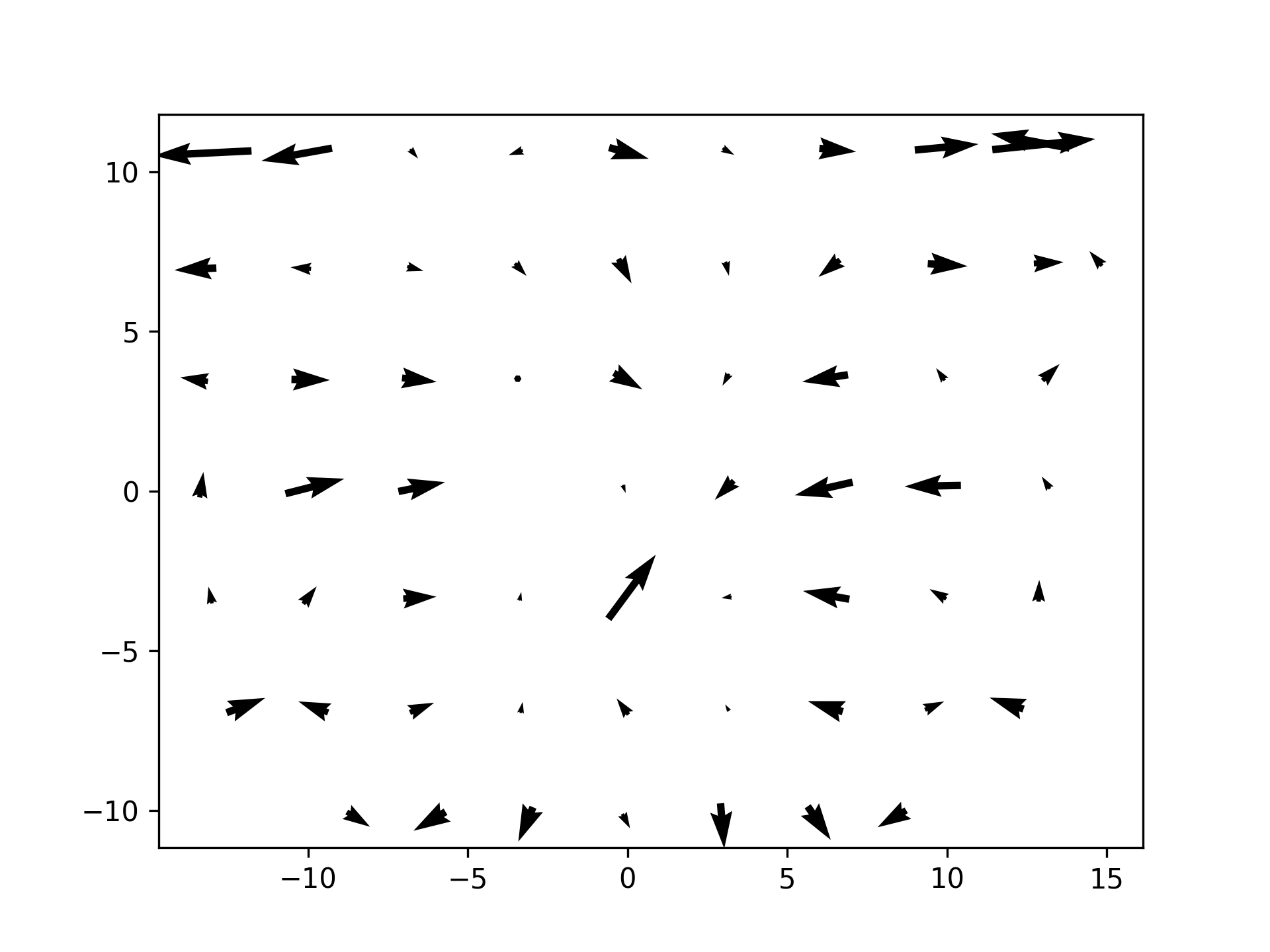}}
    \caption{Visualisation of the projection function. The arrows show the difference between the found peaks on the heat map and the results calculated form our mapping model. The arrows were magnified by a factor of three. It turns out, that the differences are clearly in pixel order.}
    \label{fig:arrows}
\end{figure}

We have found genetic algorithms to be optimal for determining the parameters of the projection function. A genetic algorithm is a metaheuristic method based on the principles of genetics and natural selection \citep{holland1975}. The problem was modeled in such a way, that each candidate solution (from now on: individual) consisted of a set of the nine parameters: $K_1,a_{00},b_{00},a_{10},b_{01}$, $a_{12}$, and the small rotation angles, $\alpha$, $\beta$ and $\gamma$. The $K_1$ parameter was first fixed by determining the parameters for the first sensor, and then the same value was used for all the other sensors. The fitness of an individual was calculated as the inverse of the sum of the differences between the measured positions and the ones reconstructed using the parameters of the individual. Our algorithm employed mutation, elitist selection, and a three-point crossover. The genetic algorithm converged for 10$\,$000 generations of 300 individuals. Other parameters were determined for each sensor independently. The errors of the parameters were derived by repeating the measurements for each sensor 100 times and then calculating the standard deviation of the parameters for each set of measurements. Fig \ref{pic:peaks} shows an example how the 100 identified peak positions scatter at a fixed pitch-yaw value. Table \ref{tab:parameters} contains the determined parameters and their errors for each sensor and Fig \ref{fig:arrows} visualises the results of our model. The deviations between the values of a given parameter for the different sensors arise from the manufacturing process.

\begin{table}[!ht]
    \centering
    \begin{tabular}{c | c c c c c }
        \hline
        \hline
         & $a_{00}$ & $b_{00}$ & $a_{10}$ & $b_{01}$ & $a_{12}$ \\
        \hline
        0x21 & $-0.78 \pm 0.10$ & $1.65 \pm 0.52 $ & $19.61 \pm 0.14 $ & $19.17 \pm 0.09$ & $-4.14 \pm 0.16$ \\
        0x22 & $-0.42 \pm 0.13$ & $2.47 \pm 0.52 $ & $19.24 \pm 0.14 $ & $19.02 \pm 0.35$ & $-3.85 \pm 0.39$ \\
        0x23 & $-0.10 \pm 0.08$ & $1.85 \pm 0.62 $ & $18.95 \pm 0.10 $ & $18.92 \pm 0.10$ & $-4.06 \pm 0.16$ \\
        0x24 & $0.87 \pm 0.09 $ & $1.07 \pm 0.71 $ & $18.86 \pm 0.13 $ & $18.89 \pm 0.11$ & $-4.03 \pm 0.19$ \\
        0x25 & $0.01 \pm 0.16$  & $0.91 \pm 0.55 $ & $19.04 \pm 0.14 $ & $19.03 \pm 0.07$ & $-3.68 \pm 0.13$ \\
        0x26 & $-0.01 \pm 0.13$ & $1.38 \pm 0.75$ & $18.90 \pm 0.16 $ & $18.96 \pm 0.11$ & $-3.96 \pm 0.24$ \\
    \end{tabular}
    \caption{Parameters for each sensor at fixed $K_1=-0.246$. As it is expected the translation parameters $a_{00}$ and $b_{00}$ are close to zero.}
    \label{tab:parameters}
\end{table}

\subsubsection{Error estimation}
We would like to prove that neglecting the errors originating from the inaccuracy of our experimental setup is justified. Keeping only the linear order terms, the error of the $X$ heatmap coordinate is:
\begin{equation}
\mathrm{d}X_\mathrm{model}=\mathrm{d}X_\mathrm{model} ^{\mathrm{par}}+\mathrm{d}X_\mathrm{model} ^\mathrm{setup}
\end{equation}
where
\begin{equation}
\mathrm{d}X_\mathrm{model} ^{\mathrm{setup}}=a_{10} \mathrm{d} \eta + a _{12} \left( 2 \eta \xi \mathrm{d} \xi + \mathrm{d} \eta \xi ^2 \right) 
\end{equation}
\begin{equation}
\mathrm{d}X_\mathrm{model} ^{\mathrm{par}}=\mathrm{d} a_{00} + \eta \mathrm{d} a_{10} + \mathrm{d} a_{12} \eta \xi ^2
\end{equation}
Which gives us the results of:
\begin{equation}
\mathrm{d}X_\mathrm{model} ^{\mathrm{par}} \sim 0.23 \quad \quad \mathrm{d}X_\mathrm{model} ^{\mathrm{setup}} \sim 0.015
\end{equation}

The results are the same for the $Y$ coordinate. Hence, the errors from the parameters are an order of magnitude larger than those arising from the inaccuracy of the setup (i.e. errors in the measurements of the distance and position of the soldering gun ($\mathbf{P}_m$), the sizes and positions of the sensors ($\Tilde{x}$) and the radii of the bevel gears ($R_{\vartheta}$, $R_{\psi}$)). The leading order error is in the subpixel order. 

\section{Discussion and Conclusions}
\label{sec:conclusions}

We have presented an experimental setup and an analysis pipeline for determining the projection function of the MLX90640 imaging IR sensor, which is a first, crucial step in its usage for attitude determination of nano-satellites. We have chosen the MLX90640 IR sensor to conduct the experiments, however, the same tools and data analysis methods can be applied to a broad range of other sensors as well.

To determine the projection function we used a soldering gun which was detected as an infrared point source by the sensors. We determined the location of the peak on the heat map produced by the soldering gun and for error estimation this process was repeated 100 times for each angular position of the sensor. All in all 90 different angular positions were recorded for a single sensor. We used a projection function with nine independent parameters projecting the coordinates to the plane of the heat map. These parameters were determined by a genetic algorithm for each sensor.

We have found that the errors from the parameters are in the subpixel order ($\mathrm{d}X_\mathrm{model} ^{\mathrm{par}} \sim 0.23$), which means an angular equivalent distance of $\sim 40$ arcminutes. We believe that the accuracy of the projection function can be further improved in the near future by designing a more advanced calibration procedure built-in into the read-out electronics of the system in order to push down the overall accuracy.

In the present paper we did not discuss the method of detecting the horizon. This is due to the fact that it is difficult to faithfully recreate the image of the horizon in a terrestrial experiment. Although it can be observed at the bottom of the images in Figure 1, we expect the horizon to have a much sharper edge observed from space. Then simple curves might be used to fit the line of these sharp transitions, which then might be transformed to arcs using the mapping function of the sensor. With these arcs the direction of the nadir is expected to be determinable with a high accuracy. Note also that in contrary to the observation of the Sun this information is accessible during the whole orbit of the satellite and multiple sensors might be able to observe the horizon at the same time.

In the operational work of attitude determination we plan to complement the infrasensor measurement with a gyroscope in order to facilitate dead-recognition of the objects in demand (the Sun and the horizon). This integrated system is going to be tested on balloon flights in the near future. We hope that the new attitude determination method we propose will not only assist the CAMELOT project but many nano-satellite mission will benefit from these results as well.  

\begin{acknowledgement}
The authors would like to thank the support of the Hungarian Academy of Sciences via the grant KEP-7/2018, providing the financial background of our experiments. This research has been supported by the European Union, co-financed by the European Social Fund
(Research and development activities at the E\"otv\"os Lor\'and University's Campus in Szombathely, EFOP-3.6.1-16-2016-00023). We also thank the support of the GINOP-2.3.2-15-2016-00033 project which is funded by the Hungarian National Research, Development and Innovation Fund together with the European Union. 
\end{acknowledgement}

{}

\end{document}